\documentclass{JHEP3}
\usepackage{float}
\usepackage{amsthm}
\usepackage{amsmath}
\usepackage{graphicx}
\usepackage{amssymb}
\newcommand{\trace}{\text{Tr}}

\title{Black hole entropy divergence and the uncertainty principle}
\author{  Ram Brustein$^{(1,2)}$, Judy Kupferman$^{(1)}$ \\ (1)  Department of Physics, Ben-Gurion
University, Beer-Sheva, 84105 Israel \\  (2) CERN, Theory Division,
CH-1211 Geneva 23, Switzerland\\ E-mail: ramyb@bgu.ac.il,     judithku@bgu.ac.il}
\date{}
\abstract{
Black hole entropy has been shown by 't Hooft to diverge at the horizon.
The region near the horizon is in a thermal state, so entropy is linear to energy which consequently also diverges. We find a similar divergence for the energy of the reduced density matrix  of relativistic and non-relativistic field theories, extending previous results in quantum mechanics. This divergence is due to an infinitely sharp division between the observable and unobservable regions of space, and it stems from the position/momentum uncertainty relation in the same way that the momentum fluctuations of a precisely localized quantum particle diverge.  We show that when the boundary between the observable and unobservable regions is smoothed the divergence is tamed. We argue that the divergence of black hole entropy can also be interpreted as a consequence of position/momentum uncertainty, and that 't Hooft's brick wall tames the divergence in the same way, by smoothing the boundary. }
\maketitle
\begin{document}

\section{Introduction}

\label{intro}

Black hole entropy diverges on the horizon. We will show that this divergence is not unique to a black hole, nor is it a UV divergence found in field theory which requires appropriate renormalization. Rather it can be seen as a result of quantum x/p uncertainty because the horizon is defined as a perfectly sharp boundary dividing spacetime into an observable and an unobservable region. A similar divergence arises for any  quantum mechanical system when a sharp boundary divides the whole system into an observed and an unobserved regions. This is also the case with a coordinate system which truncates part of flat space, as with Rindler coordinates. The divergence is tamed by smoothing over the boundary, rather than by renormalizing the theory. The same is true for black hole entropy.

In quantum mechanics we know that there are questions which can, but should not be asked. If we insist on asking them, the theory itself lets us know in a clear way by giving us a senseless answer. For example, if we ask ``what is the typical momentum of a perfectly localized particle?" the formal answer will be infinite because of the position/momentum uncertainty relation. Of course, this just means that the momentum fluctuations will become larger as the particle is localized in a sharper way. Here the observer  needs to change the question to ``what is the typical momentum of a particle whose wave function has a small finite width in space?" and treat the concept of a sharply localized particle as a limit.

In quantum field theory we are familiar with questions involving infinity. Some of these indicate real problems with divergence, but others are meaningless just as with momentum of a localized particle. An example of a real problem is to ask ``what is the charge of the electron?" where the answer comes out infinite. In this case the infinite answer does not mean that we should not have asked the question. Rather, it means that we have misidentified a microscopic parameter in the theory and that this parameter should be ``renormalized". After a redefinition of the ``bare" (correct) microscopic theory we can ask the question and get a finite answer. However, in other cases the divergence can not be corrected by modifying the theory because the question itself does not make sense.

An example of this second type of divergence would be to look at a non-relativistic particle in a finite box and ask: What is the energy in the left hand side of the box. If we approach this problem using second quantization and field operators, we will see that the problem may then be extended to relativistic fields, and that there too the difficulty arises from an ill posed question. We ask ``what is a typical energy or momentum in the left half of the box."  Note that this does not involve putting a real partition into the box; that would simply give two smaller boxes, with finite energy, of course. However if the partition is imposed by limiting the possibility of observations to only half the box without imposing new boundary conditions,  then if the partition is sharp the answer will be infinite because the fluctuations of momentum and energy are infinite.

How should we interpret the infinite answer when we know that in fact the energy is finite? In \cite{by1} it was shown that the reason for the senseless answer is that the question is inappropriate. The insistence on an infinitely sharp division between the (observable) left region and the (unobservable) right region is the cause of the divergence. In this case the sensible question should involved a smoothed division of the box, allowing the boundary between the observable and unobservable domains to be smoothed. If the resolution with which the box is divided into the observable and unobservable halves is limited, then the answer is finite and inversely proportional to the smoothing width, exactly as in the case of the localized particle.

The distinction between the two classes of divergences is the distinction between an ultraviolet (UV) divergence and an ill posed question. We would like to know to which of these two classes black hole entropy belongs. Are its divergences inherent to the system and requiring some knowledge of the UV properties of the theory, a theory of quantum gravity, or both? Or are they, rather, similar to those one obtains when dividing space into two regions, one observable and the other unobservable, and tracing over the unobservable region?

In this paper we will consider some generic wave function and apply a ``window-function" to it, leaving the boundary conditions exactly the same as they were initially. The window function will allow us to impose a smooth division between the observable and unobservable regions. When the width of the window function is taken to zero, a sharp division between the regions is obtained. This set up is different than setting up the quantum system with boundary conditions that would have made the wave function vanish outside a certain region.

We will show that the
problem of divergence at the dividing boundary can be resolved for a quantum mechanical system by asking the right question, namely smoothing the division between the two regions. We will then argue that the origin of the divergences encountered for black holes is similar. This will allow us to argue that such divergences are not a unique black hole characteristic but rather a result of quantum uncertainty, and the correct expression must involve smearing out the boundary. In fact Bekenstein noted in 1994  that if the boundary of the region being traced out were absolutely sharp, the energy would be very large due to the uncertainty principle, and so the boundary must be thought of as "slightly fuzzy" \cite{Bekenstein}, and we will show in detail that this is the case.

This paper is organized as follows. First we briefly review black hole entropy, focusing on behavior at the horizon. Then we show the relationship between entropy and  energy near the horizon of a black hole. Next, we clarify the concept of partitioning and define an operator which may smooth a partition. This is then used to examine behavior of energy at a boundary between two subsystems, first for the non-relativistic
and then for the relativistic case, and to show that in both cases energy
diverges as the boundary becomes sharp. We extend this to Rindler space, as a partitioning of Minkowsky space. Finally, we examine 't Hooft's calculation of black hole entropy, and find that his
relocation of the boundary to avoid divergence is equivalent to smearing out the boundary. Therefore here too the divergence is related to
sharpness of the boundary and is not unique to a black hole.

\section{Entropy at the horizon}

\label{review}

Black hole entropy has been calculated in various ways. These include treating it as entropy of a thermodynamic system, and as entanglement entropy which characterizes quantum correlations between two subsystems. It turns out that for an equilibrium black hole these two coincide, and its entropy is proportional to energy.

A quantum system is entangled if it cannot be expressed as a tensor product of its subsystems. In this case although the total state is pure its subsystems are in a mixed state. Entanglement entropy quantifies the extent to which a state is mixed: $S\left(\rho\right)=-Tr\left(\rho \ln\rho\right)$,
where $\rho$ is the reduced density matrix of either of the subsystems \cite{Horodecki}. If the universe is described as a pure state, the
black hole horizon divides it into that within and that without the
hole, each of which is a mixed state. Therefore black holes have entanglement entropy by definition, and the question is whether black hole entropy is anything more, or whether entanglement entropy saturates the definition.

't Hooft calculated thermodynamic characteristics of a black  hole, among them entropy, and in doing so found a divergence of the  density of states and hence of the entropy density at the horizon \cite{t'Hooft}. He overcame the problem by  adjusting the limits of integration to a brick wall a finite  infinitesimal distance from the horizon. Entanglement entropy also diverges, but the divergence appears to be an ultraviolet divergence that does not seem to diverge at any particular location.

For a BH in equilibrium, the space just outside the hole near the horizon can be treated as a thermal state in Rindler space \cite{KabatStrassler,BrusteinEinhornYarom}. In this case entanglement entropy coincides with thermal entropy, as follows. To find entanglement entropy we take the trace of part of the system. If that part of the system is a thermal state, the partial trace is a thermal density matrix,
\begin{eqnarray}
\rho_{part} & = & \frac{1}{Z}\sum_{i}e^{-\beta  E_{i}}|E_{i}\rangle\langle E_{i}|.
\end{eqnarray}
Entanglement entropy is given by
\begin{eqnarray}
S & = & -Tr\left(\rho_{part}\ln\rho_{part}\right)
\end{eqnarray}
and the energy is given by
\begin{eqnarray}
\left\langle E\right\rangle  & = & \frac{1}{Z}\sum_{i}E_{i}e^{-\beta E_{i}}
\end{eqnarray}
It follows that
\begin{eqnarray}
S & = & -\frac{1}{Z}\sum_{i}e^{-\beta E_{i}}\times\left(-\beta\sum_{i}E_{i}-\ln Z\right)\nonumber \\
& = & \beta\langle E\rangle+\ln Z.
\end{eqnarray}
For a scalar field  at a finite temperature $\ln Z$ is a constant,
so  the entropy is linear to the expectation value of the energy. Therefore in the case of a  black hole the entanglement entropy behaves as does the energy. Thus instead of examining entropy at a
barrier dividing the two subsystems, which is a complicated non-local quantity, we can calculate the reduced density matrix of a subsystem and look at the behavior of its energy which is a simpler local quantity.

\section{Momentum fluctuations, energy and entropy for smooth partitions}

\label{fluctuations}

\subsection{Partitioning a subvolume}

We will examine various examples of partitioning: first we take a single non relativistic particle in a box, then a relativistic field, then an entire region of Minkowsky space and finally a black hole. The first case is clearest but our claim is that the others are essentially the same.

It is crucial to clarify that the partitioning corresponds to limiting the observability to a subvolume. If we were to take a box and place an actual physical partition in the middle, this would impose new boundary conditions and we would simply have two smaller boxes with observables appropriate to the new boundary conditions. Instead we leave the particle in the original box, but consider only a subvolume of the box. An example of this would be to work out the probability of finding the particle. Had we actually partitioned the box and looked for the particle in the left half, we would find a probability of one or zero to find it there. But we do not actually do this; rather than making the actual observation, we just calculate the probability to find the particle on the left, and then we will obtain a probability of one half. Similarly in what follows we will calculate expectation values for part of a system without actually imposing a partition with new boundary values.

This kind of partitioning is equivalent to tracing out part of the system. The mathematical operation of tracing defines in a clear way the kind of partitioning of the quantum system that we have in mind.  We do not impose new boundary conditions, but rather we restrict the domain of observability to a limited region of the total volume. This will be implemented by a window operator, as described below. If the partitioning is done at a sharply localized point we will see divergence of momentum and energy, even though in fact obviously the particle itself has the same finite energy it had initially. If the partition is not sharply localized we will no longer see a divergence.

We are interested in the expectation value for the reduced energy in the case where we look at a subvolume of the entire system. This can be expressed in two ways. We can rewrite the state so that it is multiplied by a window function: $\left|\psi\right\rangle_{window}=f(\vec{r},w)\left|\psi\right\rangle $. Thus the expectation value for the reduced energy in this restricted system will be $\left\langle \psi\right|fHf\left|\psi\right\rangle $. Rather than regarding the window function as part of the state, we can treat as part of the operator, so that we define the restricted Hamiltonian as $H^V=fHf$. A striking equation relates quantum expectation values of operators that act on part of a system to the statistical averages for a reduced density matrix of the subsystem. Writing the density matrix for the subsystem as $\rho^V$,
\begin{equation}
\langle\psi| H^V|\psi\rangle=\trace\left(\rho^V H^V\right).
\end{equation}
Therefore we can calculate the reduced energy in the subsystem by taking the expectation value of the restricted Hamiltonian in the entire system.

The restricted Hamiltonian may be smoothed so that the partition into subsystems is not completely sharp. This is equivalent to giving the window function varying width. For details please see appendix~\ref{appA}. It is possible to define a smoothing  function that is strictly zero on the left and continuous at any fixed desired order at the boundary.  When we discuss the horizon in Rindler and Schwartzschild metrics, we will see that this is the form which can be given to a smoothing function operating on the redshift.

The function $f(\vec{r},w)$ behaves as a window enclosing part of space, and thus it mimics the
horizon by {}``truncating'' part of space for the field.
We provide it with a varying width, and examine energy as a function of its width. Our aim is to see how sharp localization affects
the reduced energy divergence.

\subsection{Energy and momentum fluctuations in a restricted non relativistic system}

We now write the reduced density matrix for nonrelativistic bosons restricted (in the sense defined above) to one part of space. We emphasize again: this restriction is related to limiting the region in which observations can be made, without imposing new boundary conditions. In practical terms it could mean adding a Heaviside step function as our window operator, thus integrating only up to a defined point. We calculate the energy we as observers will measure. We take free spinless bosons and consider states that are created by the field operator $\Psi$ acting on the vacuum:
\begin{eqnarray}
\left|\psi\right\rangle  & = & \Psi^{\dagger}(\vec{r})\left|0\right\rangle =\sum_{\vec{p}}\frac{e^{-i\vec{p}\vec{r}}}{\sqrt{\Omega}}g(\vec p)\, a_{\vec{p}}^{\dagger}\left|0\right\rangle.
\label{fieldop}
\end{eqnarray}
The function $g(\vec{p})$ is the wave function of the state in momentum space.\footnote{ The function $g$ will not be particularly relevant for us and in most cases we will ignore it by setting $g(\vec p)=1$. All our results can be easily generalized for the case $g(\vec p)\ne 1$.}

The Hamiltonian is given by
\begin{eqnarray}
H & = & \sum_{\vec{p}}\frac{p^{2}}{2m} a_{\vec{p}}^{\dagger}a_{\vec{p}}.
\label{eq:H}
\end{eqnarray}

The energy of a state  $|\psi\rangle$ is given by
\begin{eqnarray}
E = \left\langle\psi| H|\psi\right\rangle  & = & \left\langle 0|\Psi H\Psi^{\dagger}|0\right\rangle.
\label{eq:psiTpsi}
\end{eqnarray}
In configuration space the energy is given by
\begin{equation}
E=\intop_{-\infty}^{\infty}d^{3}r\frac{1}{2m} \left\langle 0| \nabla_{r}\Psi\left(\vec{r}\right) \nabla_{r}\Psi^{\dagger}\left(\vec{r}\right)|0\right\rangle.
\end{equation}

We calculate the energy corresponding to the restricted Hamiltonian
$E_\psi^V=\left\langle\psi| H^V|\psi\right\rangle=\trace(\rho^V H^V)$.
We replace the restricted Hamiltonian $H^V$ by its smoothed counterpart with the help of a window function $f(\vec{r},w)$, as discussed above.
Alternatively, we can use a restricted smoothed field operator (here we set $g=1$)
\begin{equation}
\label{psiv}
\Psi^V_{\text{smoothed}}=\int d^{3}r\, f\left(\vec{r}\right)\Psi^{\dagger}\left(\vec{r}\right)=\int d^{3}r\, f\left(\vec{r},w\right)\sum_{\vec{p}}\frac{e^{-i\vec{p}\vec{r}}}{\sqrt{V}}\, a_{\vec{p}}^{\dagger}=\sum_{\vec{p}} f\left(\vec{p},w\right)a_{\vec{p}}^{\dagger},
\end{equation}
with $f\left(\vec{p},w\right)$ being the Fourier transform of $f\left(\vec{r},w\right)$. Because $f\left(\vec{r},w\right)$ is a smooth function its Fourier transform suppresses large momenta and acts effectively as a high momentum cutoff.
The result of Eq.~(\ref{psiv}) is substituted into Eq.~(\ref{eq:psiTpsi}). The creation operators on
the vacuum give delta functions, resulting in
\begin{eqnarray}
E^V_{\text{smoothed} } & = &  \frac{1}{2m}\intop_{-\infty}^{\infty}d^{3}r \vec{\nabla}  f\left(\vec{r},w\right) \cdot \vec{\nabla} f(\vec{r},w)
\label{eq:nonrelativistic_energy_eqs-2}
\end{eqnarray}
In appendix~\ref{appB} we evaluate explicitly a related case, the restricted smoothed momentum squared $\left\langle\psi| (P^{2}_{\text{smooth}})^V|\psi\right\rangle$.

\begin{figure}[H]
\centering \includegraphics[scale=.8]{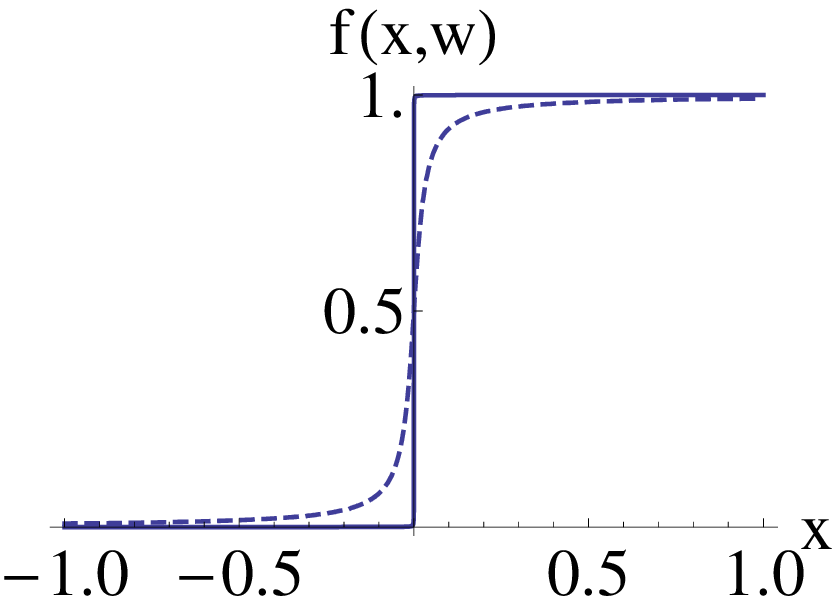} \hspace{.4in}\includegraphics[scale=.8]{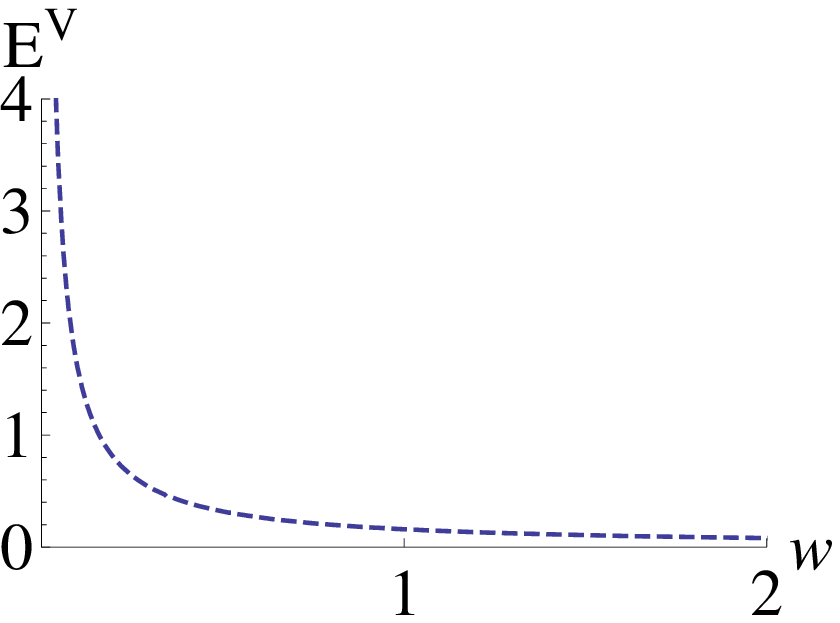} \caption{\label{fig:smoothed}Shown is a one dimensional example of a smooth window function (left) and the corresponding restricted energy as a function of barrier width (right). }
\end{figure}

For specific window functions the smoothed restricted energy can be evaluated explicitly. Consider, for example, a one dimensional case with
\begin{equation}
f(x,w)=\frac{1}{2}+\frac{1}{\pi} \arctan\left(\frac{x}{w}\right).
\end{equation}
The function is depicted by the dashed line in Fig~\ref{fig:smoothed}.
In momentum space (ignoring the singularity at $p=0$)
\begin{equation}
f(p,w)=\frac{1}{\sqrt{2\pi}}\frac{1}{p} e^{-|p|w}.
\end{equation}
So $1/w$ acts as a high momentum cutoff suppressing any momentum components of the smoothed wavefunction with $|p|>\frac{1}{w}$

The value of the restricted energy, the restriction being the positive half of the x-axis, can be calculated analytically in this case\begin{equation}
E^V_{\text{smoothed} } = \frac {1} {2m} \frac{1}{2\pi w}.
\end{equation}
This is also shown in Fig~\ref{fig:smoothed}, taking  $m=1/2$. As $w\to 0$, so that the partition becomes sharper, the energy increases, and it diverges for an infinitely sharp partition.
\footnote{ We note that this term represents the contribution of the partitioning to the energy. A full calculation would include the wave function for the particle, $g(\vec p)$ as explained in the previous note.}

Other smoothing functions yield very similar results. The restricted energy is inversely proportional to the smoothing width $w$ and diverges in the limit $w\to 0$.

For the nonrelativistic case, $ E^V_{\text{smoothed} }=\frac{1}{2m}\left\langle\psi| (P^{2}_{\text{smooth}})^V|\psi\right\rangle$. Since $\left\langle\psi| (\vec{P}_{\text{smooth}})^V|\psi\right\rangle=0$ it follows that $\left\langle\psi| (P^{2}_{\text{smooth}})^V|\psi\right\rangle=(\Delta P^V_{\text{smooth}})^2$ so the divergence of the energy is equal to the divergence of the momentum fluctuations. The divergence should not be confused with a UV divergence; the two are unrelated. The boundary behaves as if it is a localized particle. Given a function describing barrier slope, energy increases as the barrier grows sharper. That is, the more sharply the position of the dividing barrier is specified, the larger the energy. In the limit that the width tends to zero $w\to 0$ the energy diverges.  This is the same phenomenon found in quantum mechanical uncertainty, where the more sharply we specify the position of a particle, the greater the uncertainty of its momentum. The energy in this case is a simple function of momentum and linearly related to momentum uncertainty, so that as the momentum fluctuations diverge so will the energy. Thus the energy divergence here is an indication of position/momentum uncertainty.

\subsection{Relativistic smoothed restricted energy}

We extend the previous computation from the case of non-relativistic fields to the case of relativistic fields.  It is not immediately clear what position uncertainty means in the case of a relativistic field because the position operator is not defined in a clear way for this case.

The momentum operator, on the other hand, can be defined in a straightforward way from the energy-momentum tensor
$P_j=T_{0j}=\int \frac{d^{3}k}{\left(2\pi\right)^{3}}k_{j} a_{\vec{k}}^{\dagger}a_{\vec{k}}$
taking $c,\hbar=1$. Using the momentum operator we can have a practical definition of the uncertainty relations based on evaluation of the momentum fluctuations in a localized state corresponding to excitation of the field in a limited region of space. This is what we will use in the following, leaving the formal definitions and the deeper meaning of this definition for more philosophical discussions.
Let us consider a single particle state $\int d^3x g(\vec{x}-\vec{x}_0,w) \Psi^\dagger(\vec{x})|0\rangle$. The wavefunction of this state $g(\vec{x}-\vec{x}_0,w)$ is localized at $x=x_0$ with $w$ being the scale on which the state is spread. For example, we can take $g(x-x_0,w)\propto e^{-\frac{(x-x_0)^2}{2w^2}}$. We then  evaluate the momentum fluctuations in this state. They will grow in an inverse proportionality to the localization scale $w$ of the state. Similarly, if we have an $n$-particle state $\int \prod\limits_{1}^n d^3x_i g(\vec{x_i-x_0,w}) \prod\limits_{1}^n\Psi_j^\dagger(\vec{x_j})|0\rangle$ and we evaluate the fluctuations of the total momentum of the state, they will grow in an inverse proportionality to the localization scale $w$. Obviously, the state can have several localization scales. In that case the smallest one will be the most significant. The generalization to an arbitrary state should be clear by now.

In this context, formally, the only difference between a relativistic field and the non-relativistic field treated with second quantization is that both creation and destruction operators appear in the field operator. The formal analogy between the relativistic case and the non-relativistic one is clear and must point to a real correspondence between the two cases when considering the position/momentum uncertainty relation despite the inability to define a covariant position operator for the relativistic case.

In a relativistic system the energy operator is taken from the energy
momentum tensor: $H=T_{00}=\int\frac{d^{3}k}{\left(2\pi\right)^{3}} k_{0} a_{\vec{k}}^{\dagger}a_{\vec{k}}$.

In order to look for the various expectation values we recall the relativistic
scalar product:
\begin{equation}
\left\langle \varphi|\phi\right\rangle =- i \int d^{3}x\left[\varphi \partial_{t}\phi^*-\left(\partial_{t}\:\varphi\right)\phi^*\right]
\end{equation}
and the expression for the expectation value of the Hamiltonian in a state $|\varphi\rangle $
\begin{equation}
\left\langle \varphi\left|H\right|\varphi\right\rangle =-i\int d^{3}x\left[\varphi \partial_{t}\left(H\varphi\right)^*- \left(\partial_{t}\:\varphi\right)(H\varphi)^*\right].
\end{equation}

A smoothed state with window function, as before, can be defined as before $\int d^{3}r\, f\left(\vec{r}\right)\Psi^{\dagger}\left(\vec{r}\right)\left|0\right\rangle$ where the field operator here is the relativistic one. The resulting smoothed restricted energy is
\begin{eqnarray}
E_{\text{smooth}}^V=\left\langle\psi\left| (H_{\text{smooth}})^V\right|\psi\right\rangle  & = &
\int d^3  p\ f(\vec p,w)\ p\ f(-\vec p,w)
\nonumber \\ &=&
\int d^3 r\ f(\vec r,w)\ \sqrt{\vec \nabla^2}\ f(\vec r,w)
\end{eqnarray}
The details of the derivation are given in appendix~\ref{appC}. The result clearly has the same behavior as in the non relativistic case. Alternately, since $E^{2}\sim P^{2}$ we may calculate $\left\langle P^{2}\right\rangle $ and obtain\begin{eqnarray}
\frac{1}{2}\intop_{-\infty}^{\infty}d^{3}r \vec{\nabla} f\left(\vec{r},w\right) \cdot \vec{\nabla} f(\vec{r},w)
\label{eq:relativistic_energy sq_eqs-2}
\end{eqnarray}
This is identical to the non relativistic result, and equals $(\Delta P_{\text{smooth}}^V)^{2}$.

We saw that in the nonrelativistic treatment energy tends to diverge
the more sharply the boundary between the different parts of space is specified. The relativistic case shows the same phenomenon. Here too, the smoothing function $f(\vec r,w)$ acts as a momentum cutoff. In both cases the energy increases as the barrier width becomes narrower, and diverges for a completely sharp barrier with zero width.
In the relativistic case  $E^2 \sim P^2$ rather than $E \sim P^2$ but
we still obtain $E\sim\Delta p$ . As before, the energy is proportional to the momentum
uncertainty, and just as in the previous section, it diverges when the barrier is made sharp. This can be seen as an example of position/momentum uncertainty.

\section{Restricted energy and statistical entropy of the black hole}

\label{BH}

So far we have discussed restricted operators in flat spacetime. The restriction was implemented in an ad-hoc way by a choice of a (smoothed) theta function. In the case of the BH, spacetime is restricted in a different way. For example, in the Schwarzschild geometry $ds^2=-(1-\frac{r_s}{r}) dt^2 + \frac{1}{1-\frac{r_s}{r}}dr^2 +r^2 d\Omega^2$, the region of space inside the horizon $r<r_s$ is simply absent. So all the operators in Schwarzschild geometry are restricted operators. One can view the redshift factor $\frac{1}{1-\frac{r_s}{r}}$ as implementing the restriction by becoming infinite at the horizon $r=r_s$.

Our goal will be to explain how the redshift, acting as a restriction, creates an infinitely sharp boundary that results in divergence of the reduced energy and reduced entropy. We begin with the simpler case of Rindler spacetime, that is the spacetime of an accelerated observer in Minkowski space. Rindler space has the advantage that it is equivalent to a restriction to half of Minkowski space so this example allows us to explicitly compare the two restriction mechanisms. We will explain how we can implement the ideas of smoothing the boundary by restricting the maximal value of the redshift, and show that when smoothing is implemented all quantities are rendered finite with magnitude inversely proportional to the smoothing parameter, exactly as in the cases that we have encountered before. This will allow us to show that a similar phenomenon occurs for BH's.

\subsection{The uncertainty principle in Rindler spacetime}

We use the Minkowski space metric:
\begin{equation}
    ds^2=-dt^2+dz^2+{{d\vec{x}_{\bot}}}^2,
\end{equation}
where $z$ is the coordinate that will be used to separate space into the left and right halves $z<0$ and $z>0$ and $\vec{x}_{\bot}$ stands for the transverse coordinates. An accelerated observer whose acceleration is $a/2\pi$ lives in Rindler space whose metric is
\begin{equation}
\label{E:rindler}
    ds^2=-e^{2a\xi}d\eta^2+e^{2a\xi}d\xi^2+{{d\vec{x}_{\bot}}}^2.
\end{equation}
The Minkowski coordinates and Rindler coordinates are related by:
\begin{align}
    t(\xi,\eta)&=\frac{1}{a}e^{a\xi}\sinh a\eta\\
    z(\xi,\eta)&=\frac{1}{a}e^{a\xi}\cosh a\eta\\
    \vec{x}_{\bot}&=\vec{x}_{\bot}.
\end{align}

Choosing a fixed Rindler time, for example, $\eta=0$, we see that the $\xi$ coordinate only covers the $z>0$ half of space. The restriction is implemented by the redshift factor $e^{-a\xi}$ which diverges for $\xi\to -\infty$, corresponding to $z=0$.

As it stands, the restriction implemented by the redshift is infinitely sharp. The region $z<0$ simply does not exist in Rindler space. We wish to understand how to implement a smoothed restriction rather than an infinitely sharp one. So we analyze just how the redshift leads to divergence of $(\Delta p)^2$ and vanishing of  $(\Delta z)^2$, in order to consider how the divergence may be tamed. We consider a non-relativistic particle whose wave function has some spread $\Delta z$ in Minkowski space. For example,
\begin{equation}
\psi(z) = \frac{1}{\sqrt{2\pi (\Delta z)^2}} e^{-\frac{1}{2}\hbox{$\frac{z^2}{(\Delta z)^2}$}}.
\end{equation}
In momentum space the spread of the wave function is inversely proportional to $\Delta z$, $(\Delta p)^2 \sim 1/(\Delta z)^2$.
Viewed by an accelerated observer, the wave function at the origin $z=0$ corresponding to $\xi\to -\infty$ would be squeezed in the $\xi$ direction: $\Delta\xi= e^{ a\xi} \Delta z$. As required by the uncertainty principle the spread in momentum would increase, $\Delta p_\xi= e^{ -a\xi} \Delta p_z $. Thus finite $\Delta z$ and $\Delta p$ in Minkowsky space are adjusted by the Rindler metric, so that to the Rindler observer the position fluctuations at the origin will vanish and momentum fluctuations will diverge.

By our choice the particle is localized at the origin (any other choice would simply require a shift in the Rindler time $\eta$), so in the limit $\xi\to-\infty$ the momentum fluctuations diverge because the the wave function has been squeezed in space. This divergence obviously does not signal a breakdown of physics. It just means that considering the classical Rindler geometry when viewing a quantum particle requires closer thought. Rindler geometry imposes a restriction on Minkowski space. When the restriction is sharp, equivalent to localizing a particle at the origin, the momentum fluctuations diverge. Limiting the Rindler redshift factor tames the divergence and increases position fluctuations, thus softening the localization, and smoothing the restriction.

\subsection{Momentum fluctuations and redshift in Rindler spacetime}

In view of the previous discussion, and in preparation for the reinterpration of the 't Hooft calculation which we reviewed in Sect.~\ref{intro},
let us consider a (massless) scalar field $\phi$ that satisfies the Klein-Gordon equation
\begin{equation}
\frac{1}{\sqrt{-g}}\left( \partial_\mu \sqrt{-g} g^{\mu\nu}\partial_\nu\right) \phi =0.
\end{equation}

In Minkowski spacetime there is an exact solution to the Klein-Gordon equation. The $z$ dependent part of the solution is given by
\begin{equation}
\phi(z)= e^{\pm\hbox{$ i p z$}}.
\end{equation}
However, for the purpose of making the calculation more similar to the 't Hooft calculation we can rewrite the solution in a WKB form, where
the WKB solution is
\begin{equation}
\phi_{{ }_{WKB}}(z)= e^{\pm\hbox{$ i \int\limits^z p(z) dz$}}.
\end{equation}
Obviously, in Minkowski space $p(z)$ is a constant and the WKB solution reduces to the exact solution.
The WKB momentum can be expressed as
\begin{equation}
p^2(z)= E^2-p_\bot^2.
\end{equation}

In Rindler spacetime the WKB wave function is
\begin{equation}
\phi_{{ }_{WKB}}(\xi)= e^{\pm\hbox{$ i\int\limits^\xi  d\xi \sqrt{g_{\xi\xi}} p(\xi)$}}
\end{equation}
with
\begin{equation}
p^2(\xi)= g^{\eta\eta}E^2-p_\bot^2
\end{equation}
which is space varying.
So the WKB wave function is
\begin{equation}
\phi_{{ }_{WKB}}(\xi)= e^{\pm \hbox{$ i\int\limits^\xi  d\xi \sqrt{g_{\xi\xi}} \sqrt{g^{\eta\eta}E^2-p_\bot^2}$}}.
\end{equation}
Near the horizon $p(\xi)$ diverges as $\sqrt{g^{\eta\eta}E^2}= e^{-a\xi}E $  and the proper length $\widetilde{d\xi}=d\xi\sqrt{g_{\xi\xi}}=d\xi e^{a\xi}$ vanishes. This is a manifestation of the position/momentum uncertainty relation caused by the redshift.

Rindler space implements a sharp division of Minkowski space. That is, the Rindler observer sees a sharp cutoff at the horizon $\xi\to -\infty$. Smoothing this cutoff in momentum space means restricting the momentum $p(\xi)$ near the horizon. We saw in the previous section that restricting the redshift widens $\Delta x$ and shrinks $\Delta p$. Therefore restricting the redshift $g^{\eta\eta}$, $g^{\xi\xi}$ will smooth the cutoff.

In 't Hooft's black hole calculation the energy and entropy diverge due to a diverging density of states. In Rindler space too the density of states diverges, and we will see that that this divergence is due to the uncertainty principle.   We define the density of states near energy $E$ in Rindler space and evaluate it by counting the number of WKB solutions
\begin{eqnarray}
\pi n &=& \int d\xi e^{ a\xi} \int \frac{d^{2}p_{\bot}}{\left(2a\right)^{2}} p(\xi,E,p_\bot)
\nonumber \\ &=&
2\pi\int d\xi e^{ a\xi}\int \frac{dp_{\bot}}{\left(2a\right)^{2}}\ p_\bot \sqrt{e^{-2 a\xi} E^2-p_\bot^2}
\nonumber \\ &=&
-\frac{2}{3} \frac{\pi}{\left(2a\right)^{2}} E^{3}\int d\xi e^{-2 a\xi}
\end{eqnarray} where we have performed first the angular integral of $p_\bot$ and then the radial part. This integral diverges because of the diverging redshift factor at the horizon. So the density of states, the entropy and energy are divergent for the same reason and if the redshift factor is restricted, they all become finite.

We can smooth the partition by limiting the redshift, or alternately, by implementing a smoothing function on states of the system. This equivalent procedure will also tame the divergence. The smoothed functions that we need to count are obtained by multiplying the originl unsmoothed function by the smoothing function, $\psi(\xi)\to\psi(\xi) f(\xi,w)$, or in Fourier space $\phi(p)\to \phi(p) f(p,w)$. Recall that in momentum space the function $f(p,w)$ acted as a high momentum cutoff for $p > 1/w$.
Then for wavefunctions with energy $E$ we need to effectively restrict the Rindler momentum $p(\xi)=e^{- a\xi}\sqrt{E}$ to be $p(\xi) < 1/w$. In this context it simply means that the redshift factor is limited to some maximal value which can always be expressed as  $e^{-a\xi_{min}}$. The ``brick wall" model of 't Hooft in this context amounts to a sharp cutoff on the momentum $p(\xi)$. However, clearly, any other cutoff schemes will do the same job.
The density of states of smoothed wavefunctions is of course finite,
\begin{eqnarray}
\pi n &=& \int_{\xi_{min}} d\xi e^{ a\xi} \int \frac{d^{2}p_{\bot}}{\left(2a\right)^{2}}  p(\xi,E,p_\bot)
\nonumber \\ &=&
2\pi\int_{\xi_{min}} d\xi e^{ a\xi}\int \frac{dp_{\bot}}{\left(2a\right)^{2}}\ p_\bot \sqrt{e^{-2 a\xi} E^2-p_\bot^2}
\nonumber \\ &=&
\frac{2}{3}  \frac{\pi}{(2a)^3}E^3 e^{-2a\xi_{min}}.
\end{eqnarray}
This makes the energy and entropy finite and inversely proportional to the maximal redshift which determines the smoothing width of the division in Rindler space.

\subsection{Momentum fluctuations and entanglement entropy in Schwarzschild spacetime}

't Hooft solves the wave equation in the Schwartzschild metric,
identifies $p$, the wave number, and using a WKB approximation he
obtains the density of states. However the redshift leads
this to diverge at the horizon. The region near the black hole horizon is a thermal state in Rindler space, and indeed just as in Rindler space, limiting the redshift will prevent the divergence.

We recall the calculation in Schwarzschild coordinates. For simplicity we have chosen the scalar field to be massless. The Klein-Gordon equation in these coordinates is
\begin{equation}
\left(1-\frac{2M}{r}\right)^{-1}E^{2}\phi+ \frac{1}{r^{2}}\partial_{r}\left(r\left(r-2M\right)\partial_{r}\right)\phi -\left(\frac{l\left(l+1\right)}{r^{2}}\right)\phi=0.
\label{eq:thooft wave eq}
\end{equation}
The wave number can be defined as\footnote{This differs by a a factor $g_{rr}$ from 't Hooft's original defintion.}
\begin{equation}
p^{2}=g^{tt}E^{2}-\left(\frac{l\left(l+1\right)}{r^{2}}\right)
\end{equation}
Using a WKB approximation the density of states for a massless scalar field is given by
\begin{eqnarray}
\pi n & = &\sum\limits_{l,m} \intop_{2M} dr\sqrt{g_{rr}}\, p\left(r,l,m\right)
\label{eq:t'Hooft integral}\\
& = & \intop_{2M}dr\sqrt{g_{rr}}
\int(2l+1)dl\sqrt{g^{tt}E^{2}-  \frac{l\left(l+1\right)}{r^{2}}}\nonumber
\end{eqnarray}
where  $l,m$ are the angular parameters.
Evaluating the integral over $l$ we find
\begin{eqnarray}
\pi n & =& -\frac{2}{3} \intop_{2M} dr\sqrt{g_{rr}}r^2
\left(g^{tt}E^{2}\right)^{3/2}\nonumber \\
& = & -\frac{2}{3} E^3 \intop_{2M} dr \frac{r^2}{\left(1-\frac{2M}{r}\right)^2}
\end{eqnarray}
This integral diverges at the horizon. If we were to limit the redshift, as we did with Rindler space, there would be no divergence. Apparently 't Hooft does otherwise: he takes the lower limit a slight distance away from the
horizon, his well known {}``brick wall,'' so that the lower limit
becomes $2M+h$ . From this expression he obtains the energy and entropy,
which diverge as $h\rightarrow0$.

In fact 't Hooft's adjustment of the lower limit of the integral from $2M$
to $2M+h$ is equivalent to a change of variable which leaves the
lower limit at $2M$ but changes the redshift:
\begin{eqnarray}
\label{orint}
\intop_{2M+h} dr\left(1-\frac{2M}{r}\right)^{-2} & = & \intop_{2M} d\widetilde{r}\left(1-\frac{2M}{\widetilde{r}+h}\right)^{-2}
\end{eqnarray}
This clearly does not diverge at the horizon. The new expression is always finite and is limited by $\left(1-\frac{2M}{2M+h}\right)^{-2}\lesssim (2M/h)^2$ for $h\ll M$.

The altered
redshift is equivalent to multiplication of the original redshift in the $\widetilde{r}$ system
by a smoothing function:
\begin{eqnarray}
\left(1-\frac{2M}{\widetilde{r}+h}\right)^{-1} & = & \left(1-\frac{2M}{\widetilde{r}}\right)^{-1}\: f(\widetilde{r},h)
\end{eqnarray}
with
\begin{eqnarray}
f\left(\widetilde{r},h\right) & = & \frac{\left(\widetilde{r}+h\right)\left(\widetilde{r}-2M\right)}{\widetilde{r}\left(\widetilde{r}-2M+h\right)}.
\end{eqnarray}
Thus the change of variable implemented by the brick wall has the effect of multiplying the redshift by a smoothing function.

The original divergent integral in eq.~(\ref{orint}) can be expressed in terms of a sharp step function $\intop_{2M} dr  \left(1-\frac{2M}{{r}}\right)^{-2}=\intop_{0}dr\ \Theta(r-2M) \left(1-\frac{2M}{{r}}\right)^{-2}$. The altered integral can be expressed in terms of a smoothed step function
\begin{eqnarray}
\intop_{0}dr \left(1-\frac{2M}{r}\right)^{-2} f^2(r,h)\ \Theta(r-2M) & = & \intop_{0}dr \left(1-\frac{2M}{{r}}\right)^{-2}\widetilde{\Theta}(r-2M,h)\label{eq:soft step}
\end{eqnarray}
Thus we see that 't Hooft's changed lower limit is exactly equivalent
to smoothing the step function to a new one $\widetilde{\Theta}(r-2M,h)=f^2(r,h)\ \Theta(r-2M) $ with  width $h$. Formally the brick wall can be seen as either changing the redshift or smoothing the step function. Obviously, any other limiting procedure of the maximal redshift will render the integral finite and make the energy and entropy finite.

\begin{figure}[H]
\centering \includegraphics[scale=.8]{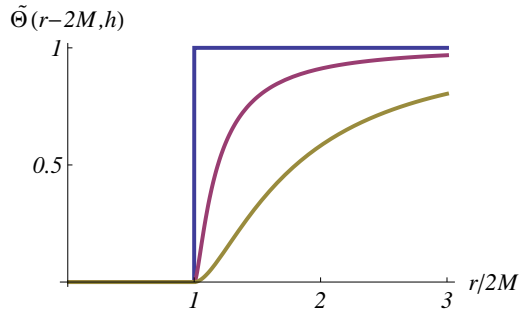}
\caption{Smoothed step function as function of $r/2M$. Curves have $h=0$  (sharp step), $h=0.1$ and $h=0.9$ (lowest) }
\end{figure}

\section{Summary and conclusions}

\label{summary and conclusions}

Energy has been shown to diverge as the boundary between two quantum subsystems, an observable subsystem and an unobservable subsystem,
becomes sharp. The divergence is due to the fact that the energy is a simple function of the momentum fluctuations. These diverge in the presence of a sharp boundary because of the uncertainty principle, much in the same way that they diverge for a sharply localized particle.  For the nonrelativistic case  $\left\langle E\right\rangle =\frac{1}{2m}(\Delta P)^{2}$. In the relativistic case $\left\langle E\right\rangle = \Delta P $  so in both cases energy divergence at an infinitely sharp boundary is clearly a consequence of position/momentum uncertainty.

In a coordinate system which implements a sharply localized boundary, the density of states and thus energy and entropy diverge at the boundary. Limiting the redshift tames this divergence. We have shown that limiting the redshift smoothes the boundary by widening $\Delta x$ and limiting $\Delta p$. Therefore the smoothing cutoff prevents the energy from diverging. This implies that the divergence of the energy and entropy was a result of the sharp localization of the boundary, and was due to the uncertainty principle.

The region near the boundary of a black hole is a thermal state, where
the entropy is linear to energy. Therefore black hole entropy will
diverge at the boundary as well. We have shown that regardless of any other cause, there would be divergence at the infinitely sharp boundary as a result of the uncertainty principle. We have also shown that 't Hooft's divergence at the black hole is an example of momentum/position uncertainty, as seen by the fact that
the {}``brick wall'' which corrects it in fact smoothes the sharp
boundary.

Our result raises the question whether the entanglement and statistical mechanics definitions of black hole entropy might refer to the same quantity. Both are proportional to area. The UV divergence may be tamed with a UV cutoff, and the boundary divergence by smearing out the boundary (both procedures might turn out to be equivalent).
So the two expressions could be expressing the same quantity.
If this is the case, then the microscopic counting of the number of
states becomes tantamount to counting the correlations between the observed and unobserved regions of spacetime. Black hole entropy has also been shown, from thermodynamic
considerations as well as explicit calculations in string
theory, to equal one fourth of the horizon area. An open problem
is to obtain the factor of $1/4$ in either definitions of black hole entropy.

\section{Acknowledgements}
We would like to thank Merav Hadad for many useful discussions. The research  was supported by The Israel Science Foundation grant no. 239/10.

\appendix

\section{Smooth restricted operators}
\label{appA}

We will now define a smoothing function which, when applied to an operator that is restricted to a sub-volume, will soften the sharp partition and serve as a momentum cutoff. Let us discuss a quantum system in a volume
$\Omega$ which is initially prepared in a pure state $|\psi\rangle$ defined in $\Omega$.  We divide the total volume into some
sub-volume $V$, and its complement $\widehat{V}$ so that $\Omega=V\oplus\widehat{V}$. The Hilbert space inherits a natural product structure ${\cal H}_{\Omega} = {\cal H}_V\otimes{\cal H}_{\widehat{V}}$.
We are interested in states $|\psi\rangle$ that are entangled with respect to the Hilbert spaces of $V$ and $\widehat{V}$ so that they can not be brought into a product form $| \psi \rangle = | \psi \rangle_V
\otimes | \psi \rangle_{\widehat{V}}$ in terms of a pure state $|\psi \rangle_V$
that belongs to the Hilbert space of $V$, and another pure state $| \psi \rangle_{\widehat{V}}$ that belongs to the Hilbert space of $\widehat{V}$.

The total density matrix is defined in terms of the total state $|\psi\rangle$
\begin{equation}
\rho=|\psi\rangle \langle\psi|.
\end{equation}
The partition of the total volume of the system into two parts
\begin{equation}
\Omega=V \oplus \widehat{V}
\end{equation}
induces a product structure on the Hilbert space and allows defining the reduced density matrix by performing a trace over part of the Hilbert space
\begin{equation}
\rho^V=\trace_{\widehat V} \rho.
\end{equation}

Operators that act on part of the Hilbert space are defined as integrals over densities in a part of space
\begin{equation}
O^V=\int\limits_V d^3r {\cal O}(\vec{r})
\end{equation}
or alternatively in terms of a theta function
\begin{equation}
\Theta^V(\vec{r})=\begin{cases} 1 & \vec{r}\in V \\ 0 & \vec{r}\in \widehat{V} \end{cases},
\end{equation}
\begin{equation}
O^V=\int\limits_\Omega d^3r {\cal O}(\vec{r})\Theta^V(\vec{r}).
\end{equation}
The relation between quantum expectation values of operators that act on part of the Hilbert space to the statistical averages with a reduced density matrix is given by
\begin{equation}
\langle\psi| O^V|\psi\rangle=\trace\left(\rho^V O^V\right).
\end{equation}

We can also define a smoothed operator
\begin{equation}
O^V_{\text {smooth}}=\int\limits_\Omega d^3r {\cal O}(\vec{r})\Theta^V_{\text {smooth}}(\vec{r},w)
\end{equation}
where $\Theta^V_{\text {smooth}}(\vec{r},w)$ represents a smoothed step function that rather than changing in a discontinuous way from zero to unity on the boundary of $V$ changes in a smooth way over a region of width $w$ near the boundary of $V$. Expressing $\Theta^V_{\text {smooth}}$ as the product of a step function and an auxiliary smoothing function $\left(f(\vec{r},w)\right)^2$ (the reason for the square will become clear in what follows):
\begin{equation}
\Theta^V_{\text {smooth}}(\vec{r},w)= \left(f(\vec{r},w)\right)^2 \Theta^V(\vec{r}) = \begin{cases} \to 1 & \vec{r}\in V\\
0 \to 1 & \vec{r}\in \partial V \hbox{ with width}\ w \\ \to  0 & \vec{r}\in \widehat{V} \end{cases}
\end{equation}
The smooth theta function defined in this way can be made continuous to any fixed desired order in derivatives. So if a class of operators has at most a given order of derivatives it is possible to define a smooth theta function that will be effectively analytic for this class.
For example, the one dimensional function
\begin{equation}
\Theta^V_{\text {smooth}}(x,w)=\begin{cases} \frac{x^n}{x^n + w^n} & x\ge 0\\
 0  & x\le 0 \end{cases}
\end{equation}
has $n-1$ continuous derivatives at $x=0$.

Rather than using the smoothed step function to modify  the operators $O^V$, we can view the smoothing function $f(\vec{r},w)$ as modifying the wave function (or state) in which the operator is being evaluated
\begin{equation}
\langle\psi| O^V_{\text {smooth}}|\psi\rangle=\langle\psi| \left(f(\vec{r},w)\right)^2 O^V |\psi\rangle = \langle f(\vec{r},w)\psi | O^V |f(\vec{r},w)\psi\rangle.
\end{equation}
Defining
\begin{equation}
|\psi_{\text {smooth}}\rangle = f(\vec{r},w)|\psi\rangle
\end{equation}
we may express the expectation value of the smoothed operator in the original state $|\psi\rangle$ in terms of an expectation value of the original operator in a smoothed state
\begin{equation}
\trace\left(\rho^V O^V_{\text {smooth}}\right)=\trace\left(\rho^V_{\text {smooth}} O^V\right)
\end{equation}
where
\begin{equation}
\rho^V_{\text {smooth}} =|\psi_{\text {smooth}}\rangle \langle\psi_{\text {smooth}}|.
\end{equation}
In momentum space
\begin{equation}
|\psi_{\text {smooth}}\rangle = \int d^3 p f(\vec p,w) \psi(\vec p) e^{-i\vec p \cdot \vec r}.
\end{equation}
Here the smoothing function $f(\vec p,w)$ looks as if it is a UV cutoff suppressing the the high momentum components of the wave function.

\section{Details of nonrelativistic smoothed momentum fluctuations}
\label{appB}

We wish to calculate  the expectation value of the smoothed operators $(P^2)^V$ which can be used to evaluate $H^V$ and other smooth operators. The partial volume $V$ is defined by a window function as described in the text.

The operator $P^2$ is given by
\begin{eqnarray}
{{P}}^{2} & = & \sum_{\vec{p}}\vec{p}a_{\vec{p}}^{\dagger}a_{\vec{p}} \cdot \sum_{\vec{k}}\vec{k}a_{\vec{k}}^{\dagger}a_{\vec{k}}\nonumber \\
& = & \sum_{\vec{p},\vec{k}}\vec{p}\cdot\vec{k}\ a_{\vec{p}}^{\dagger} \left(a_{\vec{k}}^{\dagger}a_{\vec{p}}+ \left[a_{\vec{p}},a_{\vec{k}}^{\dagger}\right]\right)a_{\vec{k}}\nonumber \\
& = & \sum_{\vec{p},\vec{k}}\vec{p}\cdot\vec{k} \left(a_{\vec{p}}^{\dagger}a_{\vec{k}}^{\dagger}a_{\vec{p}}a_{\vec{k}}+ \delta_{\vec{p}\vec{k}}a_{\vec{p}}^{\dagger}a_{\vec{k}}\right).
\nonumber \\
& = &
\sum_{\vec{p},\vec{k}}\vec{p}\cdot\vec{k} \left(a_{\vec{p}}^{\dagger}a_{\vec{k}}^{\dagger}a_{\vec{p}}a_{\vec{k}} \right)+ \sum_{\vec{p}}p^2 a_{\vec{p}}^{\dagger}a_{\vec{p}}
\end{eqnarray}
Evaluating the expectation value:
\begin{eqnarray}
& &\left\langle\psi| (P^{2}_{\text{smooth}})^V|\psi\right\rangle =  \int d^{3}r_{1}\, d^{3}r_{2}\,\left\langle 0\right|\Psi \left(\vec{r}_{1}\right)f\left(\vec{r}_{1},w\right)\, \sum_{\vec{p},\vec{k}}\vec{p}\cdot\vec{k} \left(a_{\vec{p}}^{\dagger}a_{\vec{k}}^{\dagger}a_{\vec{p}}a_{\vec{k}}+ \delta_{\vec{p}\vec{k}}a_{\vec{p}}^{\dagger}a_{\vec{k}}\right) f\left(\vec{r}_{2},w\right)\Psi^{\dagger}\left(\vec{r}_{2}\right)\left|0\right\rangle \nonumber \\
& = & \int d^{3}r\, d^{3}r_{2}\, f\left(\vec{r}_{1},w\right) f\left(\vec{r}_{2},w\right) \sum_{\vec{q},\vec{s}} \frac{e^{i\vec{q}\vec{r}_{1}}}{\sqrt{\Omega}}
\frac{e^{-i\vec{s}\vec{r_{2}}}}{\sqrt{\Omega}}
\left\langle 0\right|
\sum_{\vec{p},\vec{k}} \vec{p}\cdot\vec{k}\
a_{\vec{q}}  a_{\vec{p}}^{\dagger} a_{\vec{k}}^{\dagger}a_{\vec{p}}a_{\vec{k}}  a_{\vec{s}}^{\dagger}\,\,+
\sum_{\vec{p}} p^2 a_{\vec{q}} a_{\vec{p}}^{\dagger}a_{\vec{p}}a_{\vec{s}}^{\dagger}
\left|0\right\rangle . \hspace{.5in}
\end{eqnarray}
Since
\begin{eqnarray}
\left\langle 0\left|a_{\vec{q}}a_{\vec{p}}^{\dagger}a_{\vec{p}}a_{\vec{s}}^{\dagger} \right|0\right\rangle  & = & \delta_{\vec{p}\vec{q}} \delta_{\vec{p}\vec{s}}
\label{eq:a operators com relation}
\end{eqnarray}
and
\begin{eqnarray}
\left\langle 0\right| a_{\vec{q}}  a_{\vec{p}}^{\dagger} a_{\vec{k}}^{\dagger}a_{\vec{p}}a_{\vec{k}} a_{\vec{s}}^{\dagger}\,\,\left|0\right\rangle =0,
\label{eq:a operators com relation1}
\end{eqnarray}
the expectation value of the smooth operator is then
\begin{eqnarray}
\left\langle\psi| (P^{2}_{\text{smooth}})^V|\psi\right\rangle   & = &
\int d^{3}r\, d^{3}r_{2}\, f\left(\vec{r}_{1},w\right) f\left(\vec{r}_{2},w\right) \sum_{\vec{q},\vec{s}} \frac{e^{i\vec{q}\vec{r}_{1}}}{\sqrt{\Omega}}
\frac{e^{-i\vec{s}\vec{r_{2}}}}{\sqrt{\Omega}} p^2 \delta_{\vec{p}\vec{q}}\delta_{\vec{p}\vec{s}}  \nonumber \\
& = & \int d^{3}r\,\vec{\nabla}f\left(\vec{r},w\right) \cdot   \vec{\nabla}f\left(\vec{r},w\right) \nonumber \\
& = & \sum_{\vec{p}} p^2 f(\vec{p},w)f(-\vec{p},w)\
.
\end{eqnarray}

\section{Details of relativistic smoothed energy}
\label{appC}

In a relativistic theory the  hamiltonian is given by $\widehat H=\int\frac{d^{3}k}{\left(2\pi\right)^{3}}k_{0}a_{k}^{\dagger}a_{k}$ in momentum space. In configuration space, the expectation value of the smoothed restricted hamiltonian is given by the relativistic scalar product,
\begin{eqnarray}
\left\langle\psi\left| (H_{\text{smooth}})^V\right|\psi\right\rangle  & = & -i
\int\, d^{3}r_{1}\, d^{3}r_{2}\left[\,\,\biggl\langle 0\biggr| \Psi\left(\vec{r}_{1},t_1\right)f\left(\vec{r}_{1},w\right) \partial_{t_2}\biggl( {H}\, f\left(\vec{r}_{2},w\right)\Psi^{\dagger}\left(\vec{r}_{2},t_2\right)\biggr) -\right.
\nonumber \\
&  & \left. - \partial_{t_1}\biggl(\Psi\left(\vec{r}_{1},t_1\right) f\left(\vec{r}_{1}\right)\biggr){H}\, f\left(\vec{r}_{2}\right)\Psi^{\dagger}\left(\vec{r}_{2}\right) \biggl|0\biggr\rangle\right]_{\biggl|t_1=t_2}\equiv A-B
\end{eqnarray}
The first term $A$ is given by
 \begin{eqnarray}
A  & = & \int\, d^{3}r_{1}\, d^{3}r_{2}\, 202f\left(\vec{r}_{1},w\right) f\left(\vec{r}_{2},w\right)\biggl\langle 0\biggr|\int\frac{d^{3}p}{\sqrt{\left(2\pi\right)^{3}2p_{0}}} \left(a_{\vec{p}}e^{i\vec{p}\cdot\vec{r}_{1}-i p_0 t_1} +a_{\vec{p}}^{\dagger}e^{-i\vec{p}\cdot\vec{r}_{1}+ip_0 t_1}\right)\,\times\nonumber \\
&  &  \int\frac{d^{3}q}{\left(2\pi\right)^{3}}q_{0}a_{q}^{\dagger}a_{q} \  \times -i \partial_{t_2} \int\frac{d^{3}k}{\sqrt{\left(2\pi\right)^{3}2k_{0}}} \left(a_{\vec{k}}e^{i\vec{k}\cdot\vec{r}_{2}-ik_0 t_2}+ a_{\vec{k}}^{\dagger}e^{-i\vec{k}\cdot\vec{r}_{2}+ik_0 t_2}\right)\, \biggl|0\biggr\rangle_{\biggl|t_1,t_2=0},
 \label{Aeq1}
 \end{eqnarray}
where $p_0^2=\vec p^2$, $k_0^2=\vec k^2$.  The second term $B$ can be expressed in a similar straightforward manner.

We first perform the momentum integrals and evaluate the expectation value.
This integral includes the following sets of operators:
\[
a_{\vec{p}}a_{q}^{\dagger}a_{q}a_{\vec{k}}\,,\, a_{\vec{p}}a_{q}^{\dagger}a_{q}a_{\vec{k}}^{\dagger}\,,\, a_{\vec{p}}^{\dagger}a_{q}^{\dagger}a_{q}a_{\vec{k}}\:,\, a_{\vec{p}}^{\dagger}a_{q}^{\dagger}a_{q}a_{\vec{k}}^{\dagger},
\]
but only the second term yields a non-vanishing contribution,
\begin{eqnarray}
&  & \int\frac{d^{3}p} {\left(2\pi\right)^{3}} \frac{d^{3}k}{\left(2\pi\right)^{3}} \ d^{3}q \frac{q_0 k_0}{\sqrt{2 k_{0}}\sqrt{2 p_{0}}}\biggl\langle 0\biggr|\left(a_{\vec{p}}e^{i\vec{p}\cdot\vec{r}_{1}}+a_{\vec{p}}^{\dagger}  e^{-i\vec{p}\cdot\vec{r}_{1}}\right)\,  a_{q}^{\dagger}a_{q}\left(a_{\vec{k}}e^{i\vec{k}\cdot\vec{r}_{2}}  +a_{\vec{k}}^{\dagger}e^{-i\vec{k}\cdot\vec{r}_{2}}\right)\biggl|0\biggr\rangle
\nonumber \\ &=& \int\frac{d^{3}p} {\left(2\pi\right)^{3}} \frac{d^{3}k}{\left(2\pi\right)^{3}} \ d^{3}q \frac{q_0 k_0}{\sqrt{2 k_{0}}\sqrt{2 p_{0}}} e^{i\vec{p}\cdot\vec{r}_{1}-i\vec{k}\cdot\vec{r}_{2}} \biggl\langle 0\biggr| a_{\vec{p}}\ a_{q}^{\dagger}\ a_{q}\ a_{\vec{k}}^{\dagger} \biggl|0\biggr\rangle
\nonumber \\ &=& \int\frac{d^{3}p} {\left(2\pi\right)^{3}} \frac{d^{3}k}{\left(2\pi\right)^{3}} \ d^{3}q \frac{q_0 k_0}{\sqrt{2 k_{0}}\sqrt{2 p_{0}}} e^{i\vec{p}\cdot\vec{r}_{1}-i\vec{k}\cdot\vec{r}_{2}} \delta(\vec{p}-\vec{q})\ \delta(\vec{k}-\vec{q}).
\label{Aeq2}
\end{eqnarray}
Substituting the result of eq.~{\ref{Aeq2}} into eq.~(\ref{Aeq1}) we find
\begin{eqnarray}
A  & = & \int\, d^{3}r_{1}\, d^{3}r_{2}\, f\left(\vec{r}_{1},w\right) f\left(\vec{r}_{2},w\right)\int\frac{d^{3}p} {\left(2\pi\right)^{3}} \frac{d^{3}k}{\left(2\pi\right)^{3}} \ d^{3}q \frac{q_0 k_0}{\sqrt{2 k_{0}}\sqrt{2 p_{0}}} e^{i\vec{p}\cdot\vec{r}_{1}-i\vec{k}\cdot\vec{r}_{2}} \delta(\vec{p}-\vec{q})\ \delta(\vec{k}-\vec{q})
\nonumber \\ &=&
\int\frac{d^{3}p} {\left(2\pi\right)^{3}} \frac{d^{3}k}{\left(2\pi\right)^{3}} \ d^{3}q \frac{q_0 k_0}{\sqrt{2 k_{0}}\sqrt{2 p_{0}}} f(\vec p,w)\ f(-\vec k,w)\ \delta(\vec{p}-\vec{q})\ \delta(\vec{k}-\vec{q})
\nonumber \\ &=&
\frac{1}{2}\int d^3 p\ p\ f(\vec p,w)\ f(-\vec p,w),
\end{eqnarray}
where $p^2 =\vec{p}^2$, and $f(\vec p,w)$ is the Fourier transform of $f(\vec r,w)$.
Repeating the same steps for $B$ we find $B=-A$ so the
\begin{eqnarray}
\left\langle\psi\left| (H_{\text{smooth}})^V\right|\psi\right\rangle  & = &
\int d^3 p\ p\ f(\vec p,w)\ f(-\vec p,w)
\nonumber \\ &=&
\int d^3 r\ f(\vec r,w)\ \sqrt{\vec \nabla^2}\ f(\vec r,w)
\end{eqnarray}

\end{document}